\documentclass{article}[15pt]

\usepackage[dvips]{epsfig}
\usepackage{rotating}

\begin{document}

\title{
Spectral Analysis of Guanine and Cytosine Fluctuations
of Mouse Genomic DNA
\vspace{0.2in}
\author{
Wentian Li$^{a}$ and  Dirk Holste$^{b}$ \\
{\small \sl  a. The Robert S. Boas Center for Genomics and Human Genetics}\\
{\small \sl North Shore LIJ Institute for Medical Research, Manhasset, NY 11030, USA.}\\
{\small \sl b. Department of Biology, Massachusetts Institute of Technology, 
Cambridge, MA 02139, USA. }
}
\date{}
}
\maketitle  
\markboth{\sl W.Li, D. Holste) }{\sl W.Li, D. Holste}

{\bf key words:
DNA sequences; GC fluctuations; 1/f noise; long-ranging 
correlations; mouse genome.}

\begin{abstract}
We study global fluctuations of the guanine and cytosine base content
(GC\%) in mouse genomic DNA using spectral analyses.  Power spectra 
$S(f)$ of GC\% fluctuations in all nineteen autosomal and
two sex chromosomes are observed to have the universal
functional form $S(f) \sim 1/f^{\alpha}$ ($\alpha \approx 1$)
over several orders of magnitude in the frequency range
$10^{-7}<f< 10^{-5}$ cycle/base, corresponding to long-ranging
GC\% correlations at distances between 100 kb and 10 Mb.
$S(f)$ for higher frequencies ($f > 10^{-5}$~cycle/base) shows a
flattened power-law function with $\alpha < 1$
across all twenty-one chromosomes. The substitution of about 38\% 
interspersed
repeats does not affect the functional form of $S(f)$, indicating
that these are not predominantly responsible for the long-ranged 
multi-scale
GC\% fluctuations in mammalian genomes. Several biological
implications of the large-scale GC\% fluctuation are discussed,
including neutral evolutionary history by DNA duplication,
chromosomal bands, spatial distribution of transcription units
(genes), replication timing, and recombination hot spots.
\end{abstract}

\large

\section{Introduction}

\indent 

DNA sequences, the blueprint of almost all essential genetic
information, are polymers consisting of two complementary strands of
four types of bases: adenine (A), cytosine (C), guanine (G), and
thymine (T). Among the four bases, the presence of A on one strand is
always paired with T on the opposite strand, forming a ``base pair"
with 2 hydrogen bonds; similarly, G and C are complementary to one
another, while forming a base-pair with 3 hydrogen bonds 
\cite{pauling,watson,calladine}.
Consequently, one may characterize AT base-pairs as ``weak'' bases
and GC base-pairs as ``strong'' bases.  In addition, the frequency of A
(G) on a single strand is approximately equal to the frequency of T
(C) on the same strand, a phenomenon that has been termed ``strand
symmetry" \cite{fickett92} or ``Chargaff's second parity"
\cite{forsdyke00}. Therefore, DNA sequences can be transformed into
reduced 2-symbol sequences of weak W (A or T) and strong S (G or C) 
bases.  The
percentage of S (G or C) bases of a DNA sequence segment is denoted as
the GC base content (GC\%). 

The spatial variation of GC\% along a DNA sequence has been of
long-standing interests \cite{churchill,elton,ikemura88,ikemura90}.
GC\%-series can be considered as fluctuating or unsteady signals, and
consequently many signal processing and stochastic analysis techniques
can be applied to characterize and quantify the statistical properties
of the DNA sequences \cite{anastassiou,cristea,vaidy}.  In particular,
spectral and correlation analyzes are standard tools that can be
applied \cite{anastassiou04}.  Initial spectral
analysis \cite{li92-1,likaneko,voss} provided evidence that
DNA sequences, especially non-protein-coding sequences,
exhibit a power spectrum $S(f)$ that can be approximated by $S(f) \sim
1/f^{\alpha}$ ($\alpha \approx 1$) and are termed ``1/f noise" (or
``$1/f^\alpha$ noise'' with $ 0.5 \le \alpha \le 1.5 $)
\cite{keshner,1f,milo,west,press}.  $1/f$ noise lies in-between the
realm of white noise ($\alpha=0$) and Brownian noise ($\alpha=2$)
\cite{gardner}, and is indicative of a wide distribution of length
scales (or time, in the case of stochastic processes) \cite{peng}.

The observation of $1/f^\alpha$ spectra  in many, but not all,
short DNA sequences (of the order of a few thousands bases)
poses the question of whether $1/f^\alpha$ spectra are a
universal characteristic across all DNA sequences. Several
lines of evidence show that the $1/f^\alpha$ spectrum 
is indeed a generic phenomenon of GC\% fluctuations in DNA sequences
and is found in genomic DNA sequences from different taxonomic
classes, including genomes from bacteria \cite{maria,lu}, yeast
\cite{li-gr}, insect \cite{fukushima-t}, worm (W Li, unpublished data), 
and human \cite{fukushima,li-holste}. The human and mouse genomes
are evolutionarily separated by about 65-75 million years, and they
exhibit a high level of homology \cite{mouse}.  Yet several
species-specific differences exist that might lead to different
functional properties of $S(f)$:
\begin{itemize}
\item
While the overall, genome-wide GC\% of both human and mouse
genomic DNA sequences is about 42\%, the distribution of GC\% is
different: GC\% when measured in 20 kb ($20\times 10^{3}$ bases)
windows in mouse genomic DNA lacks extremely high and low GC\% values
\cite{mouse}.
\item
There exist pronounced differences between large sequence
segments (of the order of several Mb ($10^{6}$ bases)) of human and
mouse chromosomes due to chromosomal rearrangements. At such length
scales, GC\% correlations existing in the human genome may be absent
in mouse genome.
\end{itemize}

This Letter examines the presence of $1/f^\alpha$ spectra in
spatial GC\% variations across all {\sl Mus musculus} chromosomes.
A graphic display of the mouse genome GC\% fluctuation can
be found at \cite{paces}.

\section{Material and Methods}
\subsection{DNA Sequence Data}
We download mouse genomic DNA sequences for nineteen autosomal
chromosomes (Chr1--Chr19) and two sex chromosomes (ChrX and ChrY) from
the UCSC Genome Bioinformatics Site {\sf http://genome.ucsc.edu/}
(the October 2003 release, or UCSC version {\sl mm04}).
All twenty-one chromosomes are evenly
partitioned into $2^{17}=131,072$ non-overlapping windows of $\omega$
bases, and the GC\% of each window is computed.  A fraction of bases
are yet uncharacterized and at these positions A, C, G, or T is
substituted by the symbol ``N" (sequence gaps). Windows that
contain only uncharacterized bases have an undetermined GC\% value.
In this study, we replace all undetermined GC\% values by randomly
chosen values from a normal distribution with GC\% mean and variance
taken from the empirical distribution of all determined GC\% windows.

\begin{figure}[htbp]
  \centerline{\psfig{figure=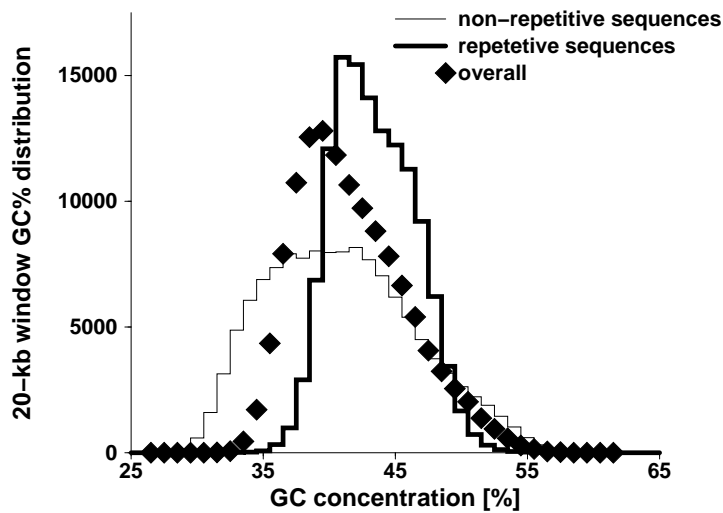,width=95mm,angle=0}}
  \caption{\label{fig1}
Genome-wide GC content (GC\%) of mouse {\em Mus musculus} 
for overall, non-repetitive, and repetitive sequences.
  }
\end{figure}

Higher eukaryotic genomes are enriched in repetitive sequences
\cite{smit}. Repeats are approximate copies of DNA sequence segments,
and interspersed repeats are an abundant class of repetitive sequence
segments in mammalian genomes that scattered throughout the
genome.  In both the human and the mouse genome, transposon-derived
interspersed repeats constitute about 35-45\%\cite{lander,venter} and
38\% \cite{mouse} of the total genome, respectively.  As can be 
seen from Fig.~1, the distribution of GC\% for repetitive 
sequences is markedly different from that of the non-repetitive sequences.
In order to study the effects of interspersed repeats,
we use interspersed repeat-annotated versions of mouse
chromosomes \cite{repeatmasker},
and separately analyze GC\% fluctuations obtained from
DNA sequences with retained and substituted interspersed repeats.  In the
latter DNA sequences, we substitute GC\% from interspersed repeats by
randomly GC\% values taken from a normal distribution with mean and
variance taken from the empirical distribution of the non-repetitive
proportion of each individual chromosome.

\subsection{Spectral analysis of DNA Sequences}

\indent

We coarse-grain sequences into a spatial-series of GC contents
(GC\%) and conduct spectral analysis of the spatial GC\%-series.  To
this end, we chose a window of size $\omega$ bases, compute GC\%, move
the window along the DNA sequence by $\Delta\omega$ bases, and iterate
the computation to obtain a spatial-series of GC\% values.
Non-overlapping windows are obtained by setting $\Delta\omega=\omega$.

The power spectrum, the absolute squared average of the Fourier
transform, is defined as
\begin{equation}
S(f) \equiv
\frac{1}{N} \left| \sum_{k=1}^{N} ({\rm GC\%})_k \cdot e^{ -i 2 \pi k 
f/N} \right|^2
\label{DEF}
\end{equation}
where $N$ is the total number of windows. Table~1 lists the window
sizes and averaged GC\% calculated at these window sizes for all
chromosomes.

\begin{table}[htbp]
\caption{\label{tab:table1}
  Window sizes ($\omega$) and average GC contents ($\overline{\rm 
GC\%}$).  Each mouse chromosome (Chr) is partitioned into $2^{17}$ 
non-overlapping windows.
  \vspace*{5mm}
  }
\centering\footnotesize
\begin{tabular}{rcc|rcc}
Chr & $\overline{\rm GC\%}$ & $\omega$~(kb) & Chr & $\overline{\rm 
GC\%}$ & $\omega$~(kb) \\ \hline
  1 & 41                & 1.52         &  11 & 44                & 0.93 \\
  2 & 42                & 1.39         &  12 & 42                & 0.88 \\
  3 & 40                & 1.25         &  13 & 42                & 0.90 \\
  4 & 42                & 1.18         &  14 & 41                & 0.90 \\
  5 & 43                & 1.15         &  15 & 42                & 0.81 \\
  6 & 41                & 1.15         &  16 & 41                & 0.76 \\
  7 & 43                & 1.05         &  17 & 43                & 0.73 \\
  8 & 42                & 1.00         &  18 & 41                & 0.69 \\
  9 & 43                & 0.96         &  19 & 43                & 0.47 \\
 10 & 41                & 1.02         &  X/Y & 39/39            & 1.21/0.17 \\
\end{tabular}
\end{table}

\section{Results}

\indent

We use the computational Fast Fourier Transform (FFT), implemented in
the {\small\protect\sf S-PLUS} statistical package (Version 3.4,
MathSoft, Inc.).  The {\small\protect\sf S-PLUS} subroutine {\small\sf
Spectrum} takes as input a discrete FFT to calculate as output a
periodogram (the power spectrum in units $10\cdot\log_{10} S(f)$), and
subsequently applies Daniell-filtering (i.e. rectangular window)
\cite{daniell,priestley} to compute a smoothed spectrum using a 
user-specified parameter value ({\sf span}).

Figure~2 shows the power spectrum $S(f)$ as a function of the
frequency $f$ across nineteen autosomal and two sex chromosomes.  We
find for sequences with retained interspersed repeats that $S(f)$
exhibits the functional form $S(f)\sim 1/f^{\alpha}$ persistently
across twenty-one chromosomes. The exponent $\alpha$ is close to $\alpha \approx$ 1
for frequency ranges of $10^{-7} < f < 10^{-5}$ cycle/base, corresponding to
length scales $L=1/f$ of 100kb  $< L <$  10Mb.  At
higher frequencies $f > 10^{-5}$ cycle/base ($L < 100$~kb), $S(f)$
generally becomes flattened with $\alpha < 1$ across all chromosomes.
This deviation from the $1/f$ spectrum was also observed in
human genome \cite{fukushima,fukushima-t,li-holste}
At lower frequencies $f < 10^{-7}$ cycle/base  ($L > 10$~Mb), there are much
less spectral components, and hence $S(f)$ shows relatively larger
fluctuations,  and the estimation of $S(f)\sim 1/f^{\alpha}$
is less reliable. In the frequency range of $10^{-7} < f < 10^{-8}$
cycle/base, only $S(f)$ for Chr2, chr4, Chr7, Chr11 and Chr16 is 
indicative of a persistence of $\alpha\approx 1$.

When we compute $S(f)$ for mouse chromosomes $1, 2, \dots, 19, X$ and Y
with substituted interspersed repeats, we find that $S(f)$ is higher than
$S(f)$ obtained for the original sequences, especially at frequency
ranges higher than $f > 10^{-5}$ ($L > 10$~kb).  One possible
explanation is that the substitution of GC\% estimated from
repetitive GC bases by random GC\% values increases the level
of white noise fluctuations at length scales comparable to lengths
of interspersed repeats.

It is interesting to note that the substitution of about 38\% interspersed
repeats hardly affects $S(f)$ at intermediate and lower frequencies. A
similar observation has been made for human genomic DNA sequences
\cite{holste03,li-holste}. Thus, interspersed repeats may not contribute
predominantly to long-ranging correlations in mammalian genomic DNA.

\begin{figure}[bh]
  \centerline{\psfig{figure=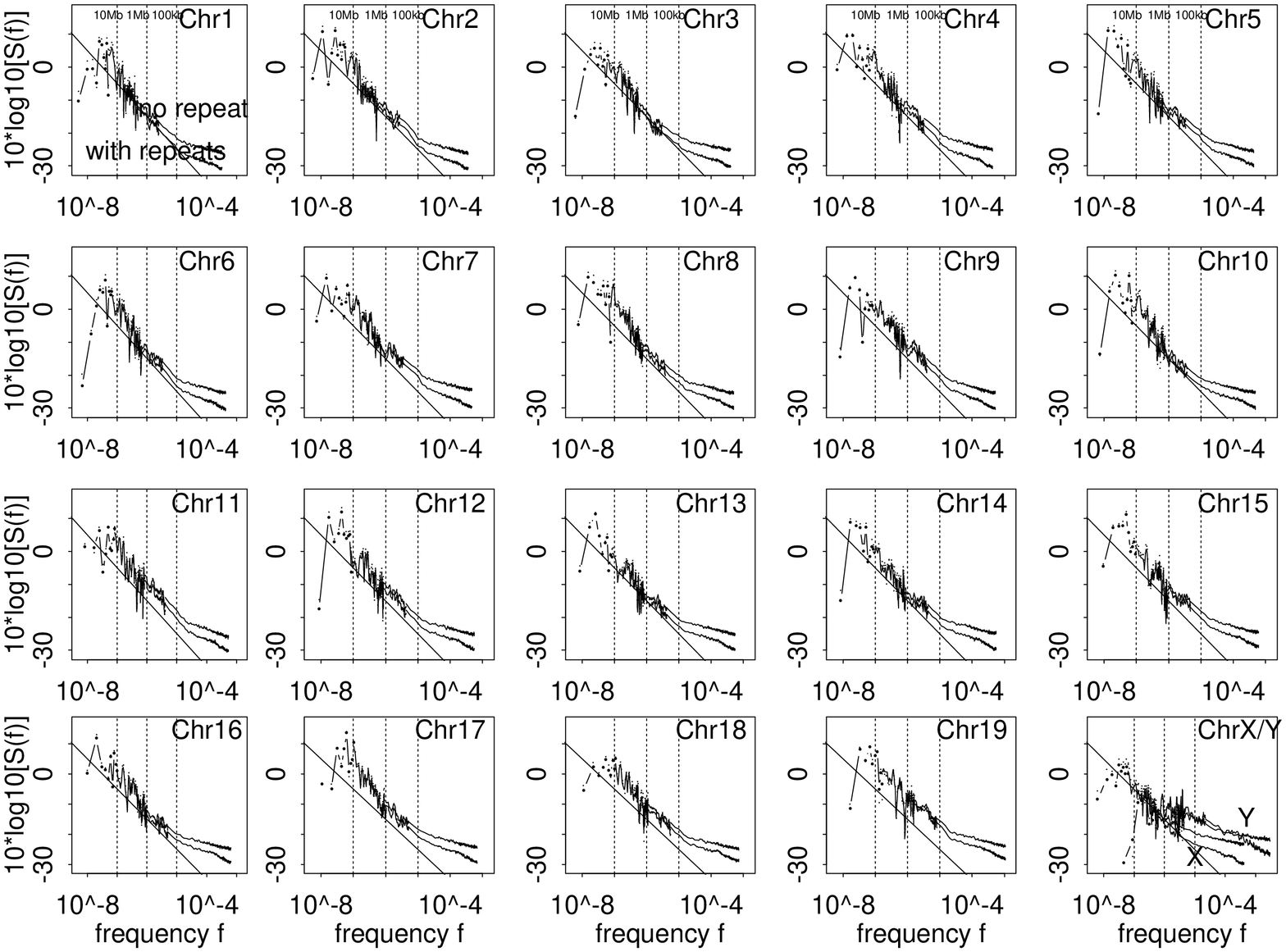,width=95mm,angle=-90}}
  \caption{\label{fig2}
  Double logarithmic representation of the power spectrum $10\log_{10}
  S(f)$ of GC\% fluctuations across nineteen autosomal (Chr1--Chr19)
  and two sex (ChrX and ChrY) mouse chromosomes.  The functional form of $S(f)$
  can be approximated by $S(f)\sim 1/f^{\alpha}$ over several order of
  frequency magnitudes ($\alpha=1$ is plotted for comparison).  Two
  curves represent $S(f)$ obtained for DNA sequences with retained and
  substituted interspersed repeats, respectively.  For clear 
representation,
  $S(f)$ is smoothed at different frequency ranges, using
  Daniell/rectangular-filter with sizes ({\sf span} parameter) 1, 3, 31,
  and 501 for the 1--10, 10--100, 100--500, and 500-65,536 spectral
  components. Vertical lines mark the length scales $L=1/f$
  (base/cycle) for $L=10$~Mb, $L=1$~Mb, and $L=100$~kb.
  }
\end{figure}

\section{Discussion}
\subsection{Spatial 1/f spectra are not an in generally held property}

\indent

Before discussing the universal spectral shape of GC\% of
genomic DNA sequences, note that not
all spatial sequences or signals exhibit $1/f^\alpha$ ($\alpha 
\approx 1$) power spectra.

An instructive example is provided  by the spatial spectrum of images 
taken of natural scenes, where it is known that the {\sl amplitude
spectrum} of image pixels is typically  $1/f$, and consequently 
its power spectrum is $S(f)\sim 1/f^2$ \cite{burton,field,tolhurst92}.
Sometimes, the exponent $\alpha$ in $S(f) \sim 1/f^\alpha$ 
may not be exactly equal to 2: for example, it was shown that 
underwater images tend to exhibit a larger
exponent of $\alpha$ (or deeper slope) than that of atmospheric
images \cite{balboa}. There are mainly two theories of the
$1/f^2$ scaling in such images: (i) it is caused by
luminance edges, and (ii) it is caused by a power-law distribution
of sizes of regions with constant intensity \cite{balboa2}.
Experiments have been carried out to test whether images with a change of the 
slope $\alpha$ can
be detected visually by human objects \cite{tolhurst97, tolhurst00}.

This well established $1/f^2$ spatial power spectrum in
images provides a case example that spatial
power spectra are not necessarily of the form $S(f) \sim 1/f$.
Rather, $S(f) \sim 1/f$ and $1/f^2$ spectra are considered to belong
to two different classes \cite{gardner}, and so the exponent
$\alpha \approx 1$ observed in DNA sequences is not 
{\sl a priori} expected.

\subsection{Spatial $1/f$ spectra are consistent with the
evolutionary expansion-modification model}

\indent

One hypothesis is that $1/f^\alpha$ ($\alpha \approx 1$) constitutes
a universal property of all long DNA sequences subject to
neutral evolution that involved duplications and
mutations \cite{li92-1,li91}. A simplified model, termed
``expansion-modification (EM) model" \cite{li89,li91} generates
a binary 2-symbol sequence by two local operations: (i)
expansion/duplication: $0 \rightarrow 00$, $ 1 \rightarrow 11$;
and (ii) modification/mutation: $0 \rightarrow 1$,
$1 \rightarrow 0$. When the probability of the first operation
is large (e.g. probability $p_1=0.9$), resulting binary sequences
exhibit $1/f^\alpha$ ($\alpha \approx 1$) power spectra.
If the probability of the second operation ($p_2$) is large, the
resulting sequences exhibit white spectra ($\alpha\approx 0$) \cite{li91}.
Since the sequence generating process is hierarchical, it
is implicit that the resulting sequence
exhibit scale-invariance (or perhaps multiple-scale-invariance
\cite{mansilla}).

The EM model contains two features that
are essential for DNA evolution: duplication and mutation.
Duplications, both inter-chromosomal
or intra-chromosomal, expand DNA sequences and provide a
potential for genes to develop novel functions
\cite{ohno}. In one point of view, duplications have a larger impact
than natural selection in Darwin's evolution theory, as
duplications and the resulting redundancy actually created
the foundation upon which natural selection acts \cite{meyer}.
Although point mutations might be detrimental to 
biological fitness, they neverthless provide a potential for evolution,
perhaps on a smaller scale as compared to that for duplications.

A more realistic modeling of neutral evolution of DNA sequences
by duplication and mutation beyond the EM model is still lacking
\cite{eichler}, and
so the hypothesis that all long DNA sequences undergoing
duplications and mutations
exhibit $1/f^\alpha$ ($\alpha \approx 1$) power spectra
remains to be validated.  The results presented in this paper,
that {\sl Mus musculus} genomic DNA exhibits $\alpha \approx 1$
adds another line of evidence toward verification of this hypothesis.

\subsection{High level of chromosomal segment translocations did
not destroy $1/f$ spectra in mouse genomic DNA}

\indent

About 90\% of the human and 93\% of the mouse
genome reside within syntenic blocks, in which the
order of a series of biological markers (e.g. genes) are
approximately conserved \cite{mouse}. However, with the
exception of ChrX, these syntenic blocks have different loci
at human and mouse chromosomes.
About 65-75 million years ago, the human genome and the
mouse genome embarked on a different evolutionary history,
with many chromosomal translocations, that left
syntenic blocks of a human chromosome
scattered on different mouse chromosomes.

This basic picture indicates that the observation of $1/f$
spectra in the human genome \cite{li-holste} will not guarantee
similar spectra in mouse genome, as random translocations
can easily destroy any long-range order a given DNA sequence.
An alternative explanation of $1/f$ spectra in the mouse
genome, in light of observed $1/f$ spectra in human genome,
is that both are in fact a consequence of the large-scale
dynamics, such as translocation and duplication. Theoretically,
DNA sequences frozen at about 65-75 million years ago before
the divergence of human and mouse species could test this
hypothesis.

Besides translocation/duplication, mutations may also affect
GC\% fluctuations. If chromosomal segment translocations
were the only process in the evolution, the human and mouse genomes
ought to have the same GC content distribution.
Nevertheless, the GC\% distribution in the mouse genome is tighter
(with smaller variance) and lacks segments with higher GC\%
(``outliers'') as in the human genome, a fact both observed
experimentally \cite{thiery} and by sequence analysis \cite{mouse}.

One might think that this difference of variances of
the GC\% distributions between the human and the mouse 
genome may render their spatial spectra different. However, 
as pointed out in \cite{clay,li04},
the exponent $\alpha$ in $1/f^\alpha$ is related to how
the variance of GC\% changes with the window size,
instead of the variance itself (and the isochore
or fairly homogeneous sequence segment of about 200--300 kb
long or more \cite{bernardi89,bernardi01} is related to
the asymmetry or the third moment of the GC\% distribution).

\subsection{GC content correlates with many biological features
and $1/f$ spectra of GC\% imply a scale-invariant distribution
of these features}

\indent

Although variations in GC\% are not of direct biological relevance,
they are correlated with many other other measurements and
biological functions of DNA sequences, including chromosome
bands, protein-coding gene density, replication timing,
or recombination hot spots \cite{bernardi89,bernardi04}.
If the correlation between GC\% and a biological function
is strong, then the scale-invariance pattern in GC\% can be
transformed into a similar spatial structure of these 
biologically functioning units.

While a stretched out chromosome is too thin to be visible
(the diameter of a DNA thread is about $2\times 10^{-9}$m 
\cite{calladine}), during the metaphase of a cell cycle 
it becomes visible because it is tightly packed into a chromatin 
structure (with the length of a typical human chromosome of this
compact form of about $10^{-5}$m \cite{calladine}).  Using 
Giemsa dyes to stain chromosomes leads to alternating dark and 
light bands \cite{ried}. The mechanism of this staining difference 
at different chromosomal regions is thought to be caused by the degree 
of condensation of the chromatin structure \cite{saitoh}
The connection between GC\% and Giemsa bands was long before
suggested: Giemsa-light bands (termed R-bands) are GC-rich, whereas
Giemsa-dark bands (G-bands) are GC-poor \cite{coming,ikemura88}.
This connection has been further calibrated, such that
being {\sl relatively} GC-rich or GC-poor as compared to the
flanking regions is correlated with Giemsa-bands \cite{gojobori}.
This new proposal manages to reproduce chromosome bands quite 
well by sequence analysis \cite{gojobori}.

After the sequencing of the human genome has been completed,
it was revealed that many long GC-poor regions contain a
low gene density (``gene desert") \cite{lander,bernardi01}, confirming
the earlier proposed correlation between GC\% and gene
density \cite{b85,b91,b96}. In fact, the gene density
not only increases with GC\%, but increases faster than
in a linear fashion \cite{bernardi01}. Extremely GC-rich regions
in the human genome are also extremely gene-rich, and beyond the
GC content of GC\% $\approx$ 46\% the gene-density
increases markedly.
 
Due to the large genome size for most eukaryotic species,
the replication of DNA sequences starts at multiple positions.
Specific chromosome regions replicate earlier in time, while other
regions replicate later, marked by a clear boundary between the
two types of regions.
While a correlation between replicating-timing and chromosome
bands was proposed in \cite{coming,bernardi89}
(R-band replicates earlier), the extent and biological relevance 
are still subject to investigation \cite{ikemura02}.

Furthermore, regions with high recombination rate (recombination 
ot spots) have been associated with being GC-rich \cite{clark}.
While this correlation has been established for yeast {\sl Saccharomyces 
cerevisiae}
genome sequence \cite{gerton}, no conclusive result have yet been
obtained for higher genomes such as human genomic DNA \cite{mcvean}.
The general difficulty in the determination of regional recombination rates
in human genome is that it is either indirect (using
pedigree analysis \cite{yu,kong}) or limited to only
male samples (using sperm typing \cite{jeffreys}). Even a newly
proposed population genealogy-based inference of recombination
rates \cite{mcvean} is still not a direct measurement. In addition,
it has also been proposed that recombination events increase
the local GC content \cite{birdsell,galtier,duret04}. While switching
the role of cause and effect, from a statistical (correlation) point 
of view,  the outcome remains the same. 

Instead of using the GC\% as a ``surrogate'', there also
have been attempts to
study large-scale patterns of biological units directly.
It has been shown that in circular bacterial genome sequences, 
the positions and orientations
of genes do not have any preferable length scale and are
scale-invariant \cite{audit}. This result should be directly linked
to a similar scale-invariance of GC\% in bacterial genomes.
The universally observed $1/f^\alpha$ ($\alpha \approx 1$) spectra
in the mouse {\sl Mus musculus} chromosomes, as well as in
human chromosomes \cite{li-holste}, motivates further sequence 
analysis on the spatial arrangement of functional biological units.

\section*{Acknowledgements}
W. Li acknowledges support from NIH-N01-AR12256.



\begin{thebibliography}{99}

\bibitem{pauling}
L. Pauling and R.B. Corey,
{\em A proposed structure for the nucleic acids},
{\em Proceedings of National Academy of Sciences} {\bf 39} (1953) 84--97.

\bibitem{watson}
J.D. Watson and C. Crick,
{P\em A structure for deoxyribose nucleic acid},
{\em Nature} {\bf 171}  (1953) 737--738.

\bibitem {calladine}
C.R. Calladine and H.R. Drew
{\em Understanding DNA}
(Academic Press, London, 1992).

\bibitem{fickett92}
J.W. Fickett, D.C. Torney and D.R. Wolf,
{\em Base compositional structure of genomes},
{\em Genomics} {\bf 13} (1992) 1056--1064.

\bibitem{forsdyke00}
D.R. Forsdyke and J.R.  Mortimer,
{\em Chargaff's legacy},
{\em Gene} {\bf 261} (2000) 127--137.

\bibitem{churchill}
G.A. Churchill,
{\em Stochastic models for heterogeneous DNA sequences},
{\em Bulletin of Mathematical Biology} {\bf 51} (1989) 79--94.

\bibitem{elton}
R.A. Elton,
{\em Theoretical models for heterogeneity for base composition in DNA},
{\em Journal of Theoretical Biology} {\bf  45} (1974) 533--553.

\bibitem{ikemura88}
T. Ikemura and S. Aota,
{\em Global variation in G+C content along vertebrate genome DNA:
possible correlation with chromosome band structure},
{\em Journal of Molecular Biology} {\bf 203} (1988) 1--13.

\bibitem{ikemura90}
T. Ikemura, K.N. Wada and S. Aota,
{\em Giant G+C\% mosaic structures of the human genome found by 
arrangement of GenBank
human DNA sequences according to genetic positions},
{\em Genomics} {\bf 8} (1990) 207--216.

\bibitem{anastassiou}
D. Anastassiou,
{\em Genomic signal processing},
{\em IEEE Signal Processing Magazine} {\bf 18} (2001) 8--20.

\bibitem{cristea}
P.D. Cristea,
{\em Large scale features in DNA genomic signals},
{\em Signal Processing} {\bf 83} (2003) 871--888.

\bibitem{vaidy}
P.P. Vaidyanathan and B.J. Yoon,
{\em The role of signal-processing concepts in genomics and proteomics},
{\em Journal of the Franklin Institute} {\bf 341} (2004) 111--135.

\bibitem{anastassiou04}
D. Sussillo, A. Kundaje and D. Anastassiou,
{\em Spectrogram analysis of genomics},
{\em EURASIP Journal on Applied Signal Processing} {\bf 2004} (2004) 29--42.

\bibitem{li92-1}
W. Li,
{\em Generating nontrivial long-range correlations and 1/f spectra by 
replication and mutation},
{\em International Journal of Bifurcation and Chaos} {\bf 2} (1992) 
137--154.

\bibitem{likaneko}
W. Li and K. Kaneko,
{\em Long-range correlation and partial $1/f^{\alpha}$ spectrum in a 
non-coding DNA sequence},
{\em Europhysics Letters} {\bf 17} (1992) 655--660.

\bibitem{voss}
R. Voss,
{\em Evolution of long-range fractal correlations and 1/f noise in DNA 
base sequences},
{\em Physical Review Letters} {\bf 68} (1992) 3805--3808.

\bibitem{keshner}
M.S. Keshner,
{\em 1/f noise},
{\em Proceedings of the IEEE} {\bf 70} (1982) 212--218.

\bibitem{1f}
W. Li, 
{\em An online bibliography on 1/f noise},
{\sf http://www.nslij-genetics.org/wli/1fnoise/}

\bibitem{milo}
E. Milotti,
{\em 1/f noise: a pedagogical review},
arxiv preprint, physics/0204033 (2002)
{\sf http://arxiv.org/abs/physics/0204033}.

\bibitem{west}
B.J. West and M.F. Shlesinger,
{\em The noise in natural phenomena},
{\em American Scientist} {\bf 78} (1990) 40--45.

\bibitem{press}
W. Press,
{\em Flicker noise in astronomy and elsewhere},
{\em Comments on Astrophysics} {\bf 7} (1978) 103--119.

\bibitem{gardner}
M. Gardner,
{\em Mathematical games -- white and brown music, fractal curves and 1/f 
fluctuations},
{\em Scientific American} {\bf 238} (1978) 16--32.

\bibitem{peng}
J.M. Hausdorff and C.K. Peng,
{\em Multis-scaled randomness: a possible source of 1/f noise in biology},
{\em Physical Review E} {\bf 54} (1996) 2154--2155.

\bibitem{maria}
M. de Sousa Vieira,
{\em Statistics of DNA sequences: a low frequency analysis},
{\em Physical Review E} {\bf 60} (1999) 5932--5937.

\bibitem{lu}
X. Lu, Z.R. Sun, H.M. Chen and Y.D. Li,
{\em Characterizing self-similarity in bacteria DNA sequences},
{\em Physical Review E} {\bf 58} (1998) 3578--3584.

\bibitem{li-gr}
W. Li, G. Stolovitzky, P. Bernaola-Galvan and J.L. Oliver,
{\em Compositional heterogeneity within, and uniformity between, DNA
sequences of yeast chromosomes},
{\em Genome Research} {\bf 8} (1998) 916--928.

\bibitem{fukushima-t}
A. Fukushima,
{\em Periodicity in Genome Architecture from Bacteria to Human}
(Ph.D Thesis, Nara Institute of Science and Technology, 2003).

\bibitem{fukushima}
A. Fukushima, T. Ikemura, M. Kinouchi, T. Oshima, Y. Kudo, H. Mori
and S. Kanaya,
{\em Periodicity in prokaryotic and eukaryotic genomes identified by
power spectrum analysis},
{\em Gene}, {\bf 300} (2002) 203--211.

\bibitem{li-holste}
W. Li and D. Holste,
{\em Universal 1/f noise, cross-overs of scaling exponents,
and chromosome specific patterns of GC content in
DNA sequences of the human genome}, 
{\em Physical Review E}, submitted.

\bibitem{mouse}
R.H. Waterston {\em et al.},
{\em Initial sequencing and comparative analysis of the mouse genome},
{\em Nature} {\bf 420} (2002) 520--562.

\bibitem{paces}
J. Pa\u{c}es, R. Z\'{i}ka, V. Pa\u{c}es, A. Pavl\'{i}\u{c}ek, O. Clay
and G. Bernardi,
{\em Representing GC variation along eukaryotic chromosomes},
{\em Gene} {\bf 333} (2004) 135--141.

\bibitem{smit}
A.F. Smit,
{\em Interspersed repeats and other mementos of transposable elements in 
mammalian genomes}, 
{\em Current Opinion in Genetics \& Development} {\bf 9} (1999) 657--663.

\bibitem{lander}
E.S. Lander {\em et al.},
{\em Initial sequencing and analysis of the human genome},
{\em Nature} {\bf 409} (2001) 860--921.

\bibitem{venter}
J.C. Venter {\em et al.},
{\em The sequence of the human genome},
{\em Science} {\bf 291} (2001) 1304--1351.

\bibitem{repeatmasker}
A.F. Smit and P. Green,
RepeatMasker,
{\sf http://repeatmasker.genome.washington.edu/}.


\bibitem{daniell}
P.J. Daniell,
{\em Discussion on the Paper by Bartlett, Foster, Cummingham and Hynd},
{\em Supplement to the Journal of the Royal Statistical Society} {\bf  
8} (1946) 88--90.

\bibitem{priestley}
M.B. Priestley,
{\em Spectral Analysis and Time Series}
(Academic Press, London, 1981).



\bibitem{holste03}
D. Holste, S. Beirer, P. Schieg, I. Grosse and H. Herzel,
{\em Repeats and correlations in human DNA sequences},
{\em Physical Review E} {\bf 67} (2003) 061913.

\bibitem{burton}
G.J. Burton and T.R. Moorehead,
{\em Color and spatial structure in natural scenes},
{\em Applied Optics} {\bf 26} (1987) 157--170.

\bibitem{field}
D.J. Fields,
{\em Relations between the statistics of natural images
and the response properties of cortical cells},
{\em Journal of the Optical Society of America A} {\bf 4} (1987) 2379--2394.

\bibitem{tolhurst92}
D.J. Tolhurst, Y. Tadmor and T. Chao,
{\em Amplitude spectra of natural images},
{\em Ophthalmic \& Physiological Optics} {\bf 12} (1992) 229--232.

\bibitem{balboa}
R.M. Balboa and N.M. Grzywacz,
{\em Power spectra and distribution of contrasts of natural images from 
different habitats},
{\em Vision Research} {\bf 43} (2003) 2527--2537.

\bibitem{balboa2}
R.M. Balboa, C.W. Tyler and N.M. Grzywacz,
{\em Occlusions contribute to scaling in natural images},
{\em Vision Research} {\bf 41} (2001) 955--964.

\bibitem{tolhurst97}
D.J. Tolhurst and Y. Tadmor,
{\em Discrimination of changes in the slopes of the
amplitude spectra of natural images: band-limited
contrast and psychometric functions},
{\em Perception} {\bf 26}  (1997) 1011--1025.

\bibitem{tolhurst00}
C.A. Parraga and D.J. Tolhurst,
{\em The effect of contrast randomisation on the discrimination of
changes in the slopes of the amplitude spectra of natural scenes},
{\em Perception} {\bf 29} (2000) 1101--1116.

\bibitem{li91}
W. Li,
{\em Expansion-modification systems: a model for spatial 1/f spectra},
{\em Physical Review A} {\bf 43} (1991) 5240--5260.

\bibitem{li89}
W. Li,
{\em Spatial 1/f spectra in open dynamical systems},
{\em Europhysics Letters} {\bf 10} (1989) 395--400.


\bibitem{mansilla}
R. Mansilla and G. Cocho,
{\em Multiscaling in expansion-modification systems: an
explanation for long range correlation in DNA},
{\em Complex Systems} {\bf 12} (2000) 207--240.

\bibitem{ohno}
S. Ohno,
{\em Evolution by Gene Duplication}
(Springer-Verlag, Berlin, 1970).

\bibitem{meyer}
A. Meyer and Y. van de Peer,
{\em Natural selection merely modified while redundancy created.
Susumu Ohno's idea of evolutionary importance of gene and genome 
duplication},
{\em Journal of Structural and Functional Genomics} {\bf 3} (2003) vii-ix.

\bibitem{eichler}
E.E. Eichler and D. Sankoff,
{\em Structural dynamics of eukaryotic chromosome evolution},
{\em Science} {\bf 301} (2003) 793--797.

\bibitem{thiery}
J.P. Thiery, G. Macaya and G. Bernardi,
{\em An analysis of eukaryotic genomes by density gradient centrifugation},
{\em Journal of Molecular Biology} {\bf 108} (1976) 219--235.

\bibitem{clay}
O. Clay, N. Carels, C. Douady, G. Macaya, G. Bernardi,
{\em Compositional heterogeneity within and among isochores in mammalian 
genomes},
{\em Gene} {\bf 27} (2001) 615--24.

\bibitem{li04}
W. Li,
{\em Large-scale fluctuation of guanine and cytosine content
in genome sequences: isochores and 1/f spectra},
in: {\em Progress in Bioinformatics} (Nova Science, 2005).

\bibitem{bernardi89}
G. Bernardi,
{\em The isochore organization of the human genome},
{\em Annual Review of Genetics} {\bf 23} (1989) 637--661.

\bibitem{bernardi04}
G. Bernardi, {\em Structural and Evolutionary Genomics}
(Elsevier, 2004).


\bibitem{bernardi01}
G. Bernardi,
{\em Misunderstandings about isochores. Part I},
{\em Gene} {\bf 276} (2001) 3--13.

\bibitem{ried}
T. Ried,
{\em Cytogenetics -- in color and digitized},
{\em New England Journal of Medicine} {\bf 350} (2004) 1597--1600.

\bibitem{saitoh}
Y. Saitoh and U.K. Laemmli,
{\em From the chromosomal loops and the scaffold to the classic
bands of metaphase chromosomes},
{\em Cold Spring Harbor Symposium on Quantitative Biology} {\bf 58} 
(1993) 755--765.

\bibitem{coming}
D.E. Comings,
{\em Mechanisms of chromosome banding and implications for chromosome 
structure},
{\em Annual Review of Genetics} {\bf 12} (1978) 25--46.

\bibitem{gojobori}
Y. Niimura and T. Gojobori,
{\em {\rm In silico} chromosome staining: reconstruction
of Giemsa bands from the whole human genome sequence},
{\em Proceedings of the National Academy of Sciences} {\bf 99} (2002) 
797--802.

\bibitem{b85}
G. Bernardi, B. Olofsson, J. Filipski, M. Zerial, J. Salinas,
G. Cuny, M. Meunier-Rotival and F. Rodier,
{\em The mosaic genome of warm-blooded vertebrates},
{\em Science} {\bf 228} (1985) 953--958.

\bibitem{b91}
D. Mouchiroud, G. D'Onofrio, B. Aissani, G. Macaya, C. Gautier
and G. Bernardi,
{\em The distribution of genes in the human genome},
{\em Gene} {\bf 100} (1991) 181--187.

\bibitem{b96}
S. Zoubak, O. Clay and G. Bernardi,
{\em The gene distribution of the human genome},
{\em Gene} {\bf 174} (1996) 95--102.

\bibitem{ikemura02}
Y. Watanabe, A. Fujiyama, Y. Ichiba, M. Hattori,
T. Yada, Y. Sakaki and T. Ikemura,
{\em Chromosome-wide assessment of replication timing for
human chromosomes 11q and 21q: disease-related genes in timing-switch 
regions},
{\em Human Molecular Genetics} {\bf 11} (2002) 13--21.

\bibitem{clark}
S.M. Fullerton, A. Bernardo Carvalho and A.G. Clark,
{\em Local rates of recombination are positively correlated with GC 
content in the human genome},
{\em Molecular Biology and Evolution} {\bf 18} (2001) 1139--1142.

\bibitem{gerton}
J.L. Gerton, J. DeRisi, R. Shroff, M. Lichten, P.O. Brown and T.D. Petes,
{\em Global mapping of meiotic recombination hotspots
and coldspots in the yeast {\rm Saccharomyces cerevisiae}},
{\em Proceedings of the National Academy of Sciences} {\bf 7} (2000) 
11383--11390.

\bibitem{mcvean}
G.A.T. McVean, S.R. Myers, S. Hunt, P. Deloukas, D.R. Bentley and P. Donnelly,
{\em The fine-scale structure of recombination rate
variation in the human genome},
{\em Science} {\bf 304} (2004) 581--584.

\bibitem{yu}
A. Yu, C. Zhao, Y. Fan, W. Jang, A.J. Mungall, P. Deloukas, A. Olsen,
N.A. Doggett, N. Ghebranious, K.W. Broman and J.L. Weber,
{\em Comparison of human genetic and sequence-based physical maps},
{\em Nature} {\bf 409} (2001) 951--953.

\bibitem{kong}
A. Kong, {\em et al.}
{\em A high-resolution recombination map of the human genome},
{\em Nature Genetics} {\bf 31} (2002) 241--247.

\bibitem{jeffreys}
A.J. Jeffreys, A. Ritchie and R. Neumann,
{\em High resolution analysis of haplotype diversity and meiotic crossover
in the human TAP2 recombination hotspot},
{\em Human Molecular Genetics} {\bf 9} (2000) 725--733.

\bibitem{birdsell}
J.A. Birdsell,
{\em Integrating genomics, bioinformatics, and classical genetics to
study the effects of recombination on genome evolution},
{\em Molecular Biology and Evolution} {\bf 19} (2002) 1181--1197.

\bibitem{galtier}
J.I. Montoya-Burgos, P. Boursot and N. Galtier,
{\em Recombination explains isochores in mammalian genomes},
{\em Trends in Genetics} {\bf 19} (2003) 128--130.


\bibitem{duret04}
J. Meunier and L. Duret,
{\em Recombination drives the evolution of GC-content in the human genome},
{\em Molecular Biology and Evolution} {\bf 21} (2004) 984--990.

\bibitem{audit}
B. Audit and C.A. Ouzounis,
{\em From genes to genomes: universal, scale-invariant properties of 
microbial chromosome organisation}
{\em Journal of Molecular Biology} {\bf 332} (2003) 617-633.

\end{thebibliography}
\end{document}